\documentclass{icrc2009}

\usepackage{graphicx}   
\usepackage[caption=false]{caption}    
\usepackage[font=footnotesize]{subfig} 
\usepackage{fixltx2e}
\usepackage{url}

\newcommand{\shorttitle}[1]%
{\markboth{Proceedings of the 31\MakeLowercase{$^{st}$} ICRC, {\L}\'{o}d\'{z} 2009}{#1} }
\newcommand{\etal}{\MakeLowercase{\textit{et al. }}} 

\usepackage{units}

\begin{document}
\title{Sensitivity of Extensive Air Showers to Features of Hadronic Interactions at Ultra-High Energies}

\author{\IEEEauthorblockN{Ralf Ulrich,
    Ralph Engel,
    Steffen M\"uller,
    Tanguy Pierog,
    Fabian Sch\"ussler and
    Michael Unger}\\
  
  \IEEEauthorblockA{Karlsruher Institut f\"ur Technologie
    (KIT)$^1$\\ Institut f\"ur Kernphysik,
    P.O. Box 3640, 76021 Karlsruhe, Germany}}

\shorttitle{R. Ulrich \etal Interactions in Air Showers}
\maketitle

\protect\footnote{$^1$KIT is the cooperation of University Karlsruhe and
  Forschungszentrum Karlsruhe} 

\begin{abstract}
  We study the dependence of extensive air shower development on the
  first hadronic interactions at ultra-high energies occurring in the
  startup phase of the air shower cascade. The interpretation of
  standard air shower observables depends on the characteristics of
  these interactions. Thus, it is currently difficult to draw firm
  conclusions for example on the primary cosmic ray mass composition
  from the analysis of air shower data. On the other hand, a known
  primary mass composition would allow us to study hadronic
  interactions at center of mass energies well above the range that is
  accessible to accelerators measurements.
\end{abstract}


\section{Introduction}
Currently, the interpretation of existing high quality air shower data
in terms of important properties as e.g. the primary mass composition,
is complicated by the poorly constraint hadronic interaction physics
at ultra-high energies (e.g. Ref.~\cite{Antoni:2005wq}). For an
unambiguous analysis of air shower data reduced uncertainties of
interaction characteristics are needed.

The longitudinal development of extensive air showers is very
sensitive to hadronic interaction in the startup of the air shower
cascade. These few interactions at ultra-high energies are subject to
particularly large uncertainties; Their characteristics must be
inferred from extrapolations of accelerator data at much lower
energies to cosmic-ray energies and secondary particle production
phase space. These extrapolations are not well constraint by theory
nor experiment~\cite{Engel:2002id}. To explore the importance of these
extrapolations to ultra-high energies on the final resulting air
shower observables, we modified these extrapolations during air shower
simulations.

If, with the help of astrophysical arguments, the composition of
cosmic rays of a specific energy can be constraint, then it is
possible to learn about the physics of hadronic interactions at
energies far above the LHC from the analysis of air shower data. This
would allow one to use ultra-high energy cosmic ray observatories as
fixed target particle physics experiments at energies up to
$\sqrt{s}\sim\unit[450]{TeV}$, which is far above the reach of any
Earth-based particle accelerator.

\section{Modified Air Shower Simulations}
\begin{figure}[b!]
  \centerline{%
    \includegraphics[width=\linewidth]{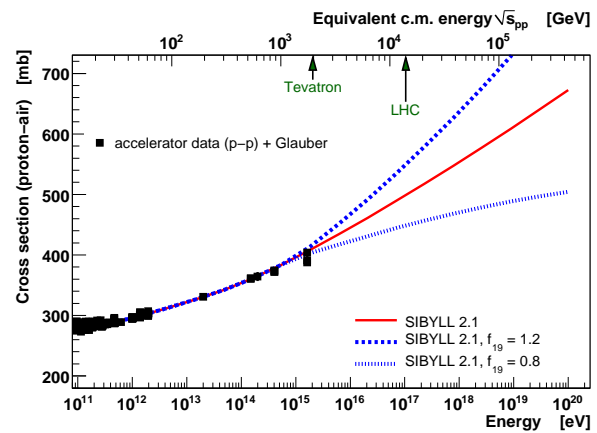}}
  \caption{Example of a modified hadronic production cross section for
    \textsc{Sibyll} for a 20\,\% increase and decrease of
    $f_{19}$~\protect\cite{Ulrich:ISVHECRI08}.}
  \label{f:SigmaModifiedCrossSection}
\end{figure}
\begin{figure*}[t!]
  \centerline{\subfloat[Multiplicity]{\includegraphics[width=.51\linewidth]{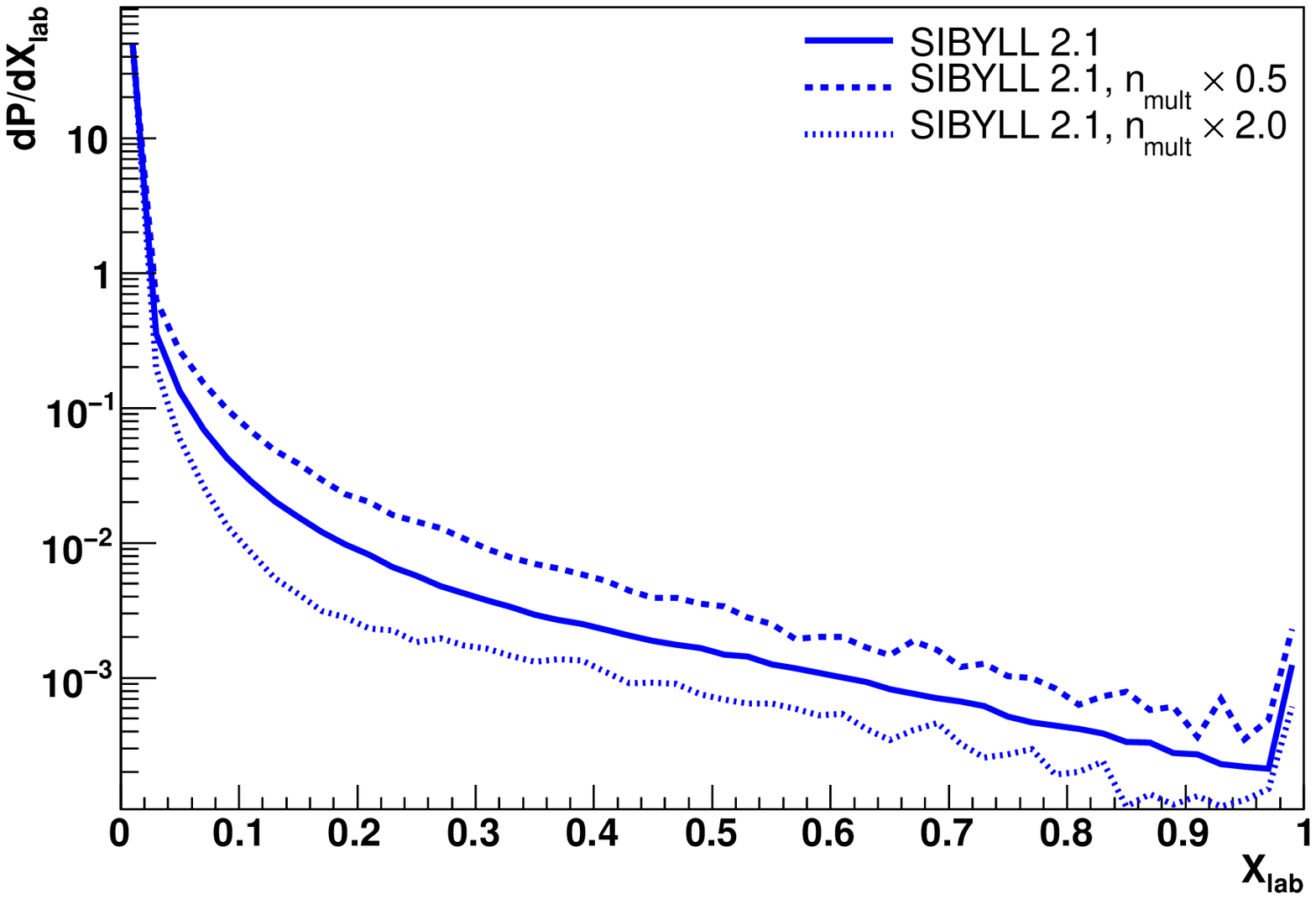}}
    \hfil
    \subfloat[Elasticity]{\includegraphics[width=.51\linewidth]{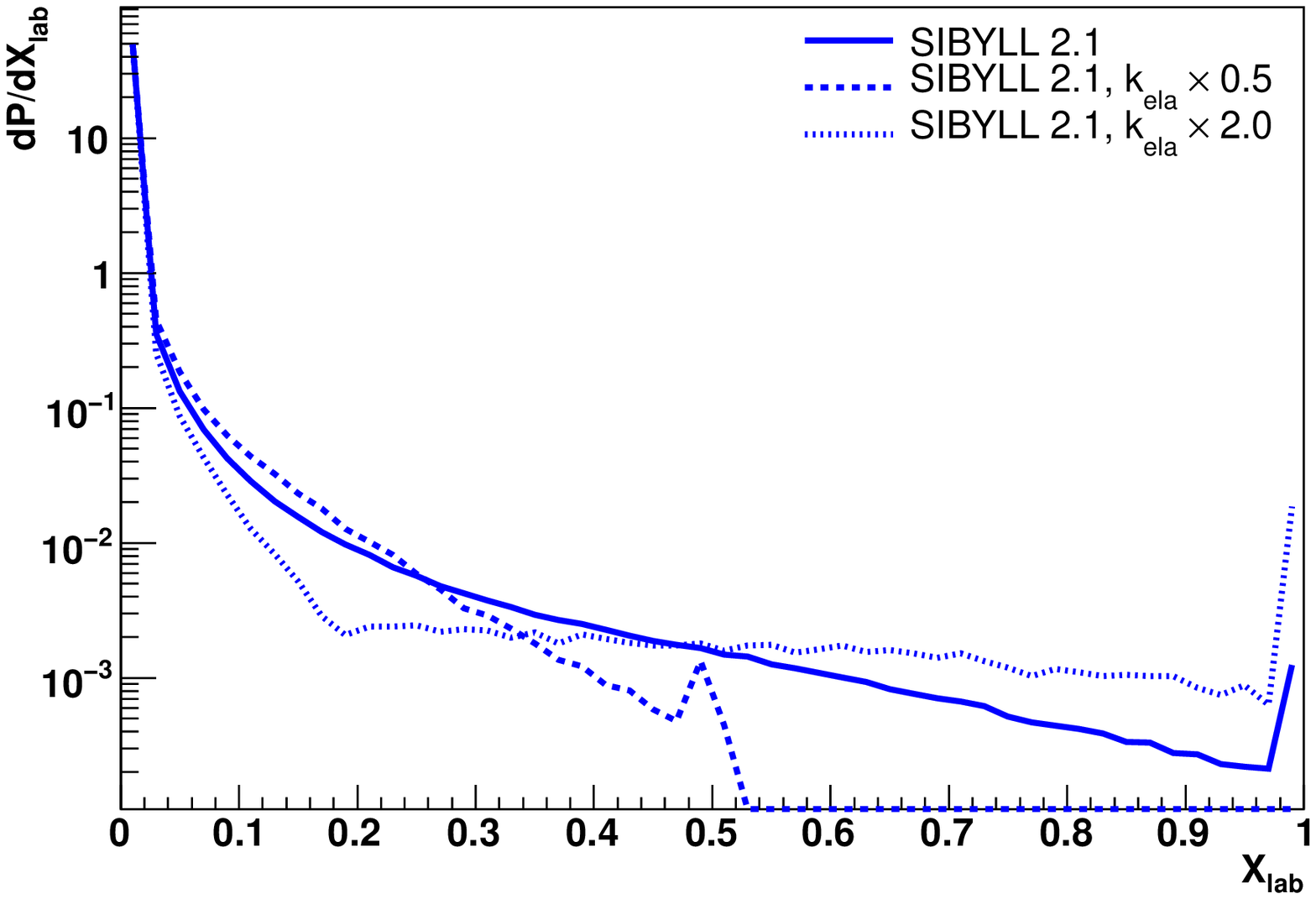}}
  }
  \caption{Impact of secondary particle resampling on $X_{\rm
      lab}$-distributions~\protect\cite{Ulrich:2009xyz}.}
  \label{fig:Xlab}
\end{figure*}
For our studies we implemented a modified version of the
\textsc{Conex}~\cite{conex} air shower simulation program that can
modify the characteristics of hadronic interactions during the
simulation. We adapt the following factor to re-scale specific
properties of hadronic interactions:
\begin{equation}
  \label{eqn:modifier}
  f(E)=1+(f_{19}-1)\; F(E) 
\end{equation}
with
\begin{equation}  
  \label{eqn:MODIFIER}
  F(E) = \left\{
  \begin{array}{l l}
    \; 0 & \quad E\le1\,\mbox{PeV}\\ \frac{\ln(E/1\,{\rm
        PeV})}{\ln(10\,{\rm EeV}/1\,{\rm PeV})} & \quad
    E>1\,{\rm PeV}
  \end{array}\right. \,,
\end{equation}
where $E$ is the energy of the projectile of the interaction.  The
factor $F(E)$ is 0 below $10^{15}\,$eV, and thus $f(E)=1$, where
accelerator data is available to constrain the models (the Tevatron
corresponds to $\sim2\times10^{15}\,$eV). At higher energies $F(E)$ is
increasing logarithmically with energy, reflecting the growing
uncertainty of the extrapolation with energy. The resulting impact of
$f(E)$ on the extrapolation of the production cross section is shown
in Fig.~\ref{f:SigmaModifiedCrossSection}. By using
Eq.~(\ref{eqn:modifier}) for all interactions during the simulated air
shower development with energies above $10^{15}\,$eV, the effect of
the modified extrapolation to ultra-high energies affects not only
the primary cosmic ray-air interaction, but also the high energy
interactions in the startup phase of the air shower, until the energy
of the particles drops below $10^{15}\,$eV.

The interactions of hadrons, and thus in particular of primary proton
cosmic ray particles, are directly modified by using the factor
Eq.~(\ref{eqn:modifier}). For nuclei, however, the semi-superposition
model~[5,$\;$6] 
is applied in order to
describe the interactions of nuclei based on the fundamental hadronic
interactions of the individual nucleons of the nucleus. Since this
model is implemented within the {\scshape Sibyll} event
generator~\cite{Fletcher:1994bd}, the handling of primary nucleons is
straightforward and all results presented here are based on the
\textsc{Sibyll} model.

To change characteristics of the secondary particle production, like
e.g. the multiplicity or the elasticity, we developed a secondary
particle re-sampling algorithm that, by deletion or duplication of
existing particles and re-distributing of kinetic energy between
secondary particles, can achieve to modify these characteristic
properties. At the same time great care is invested to conserve all
relevant physical quantities such as the total energy, leading
particles, charge, particle types and energy fractions in particle
type groups as far as possible. Also the momentum of all particles is
consistently re-calculated. A detailed description of the algorithm can
be found in Ref.~\cite{Ulrich:2009xyz}. In Fig.~\ref{fig:Xlab} the
impact of the resampling algorithm for multiplicity and elasticity on
secondary particle $X_{\rm lab}$-distributions is displayed. For the
modified multiplicity the number of particles is rescaled, while
leaving the shape of the $X_{\rm lab}$-distribution almost untouched,
in particular the leading particle is conserved. In the case of a
modified elasticity kinetic energy is re-distributed between the
leading particle and the rest of the secondaries. For an increased
elasticity the leading particle inherits energy from the other
particles, so these particles are accumulating at lower energies. For
the decreasing elasticity the leading particle looses significance by
a reduction of its energy. This leads to a generally more uniform
distribution of the total energy on all the secondaries, and in the
limiting case to the equal distribution of energy on all secondaries.

For our study we simulated 1000 air showers for each value of
$f_{19}$. All simulations were performed at primary energies of
$10^{19.5}\,$eV for proton and iron primaries.

\section{Results}
We are concentrating on three features of hadronic interactions, that
can be easily attributed a direct impact on air shower
development. These are the hadronic production cross section $\sigma$,
secondary multiplicity $n_{\rm mult}$ and the elasticity $k_{\rm
  ela}=E/E_{\rm tot}$. 
Extended Heitler models (e.g. Ref.~\cite{Matthews:2005sd}) exhibit the
relation between these quantities to air shower observable as $X_{\rm
  max}$
\begin{equation}
 \label{eqn:extendentHeitler}
    X_{\rm max} \approx \lambda_{\rm int} + \lambda_r \cdot \ln
    \frac{E_0(1-k_{\rm ela})}{n_{\rm mult}\cdot E^{\;\rm e.m.}_{\rm
        crit}}\,,
\end{equation}
 where $\lambda_r$ is the electromagnetic radiation length and
 $E^{\;\rm e.m.}_{\rm crit}$ the critical energy in air. 

To demonstrate the impact of these interaction features on the air
shower development we simulate the effect on the depth of the shower
maximum, $X_{\rm max}$, and on the total number of electrons above
$\unit[1]{MeV}$, $N_{\rm e}$, and muons above $\unit[1]{GeV}$, $N_{\rm
  \mu}$, after $\unit[1000]{g/cm^2}$ of shower development.

The quantity that is affected most directly is $X_{\rm max}$, see
Eq.~(\ref{eqn:extendentHeitler}); The effects on $N_{\rm e}$ and
$N_\mu$ can be mostly understood relative to $X_{\rm max}$ - as the
consequence of a changing distance from the shower maximum to the
observation level.

\begin{figure*}[p!]
  \centerline{%
    \subfloat[Shower maximum, $X_{\rm max}$]{\includegraphics[width=.34\linewidth]{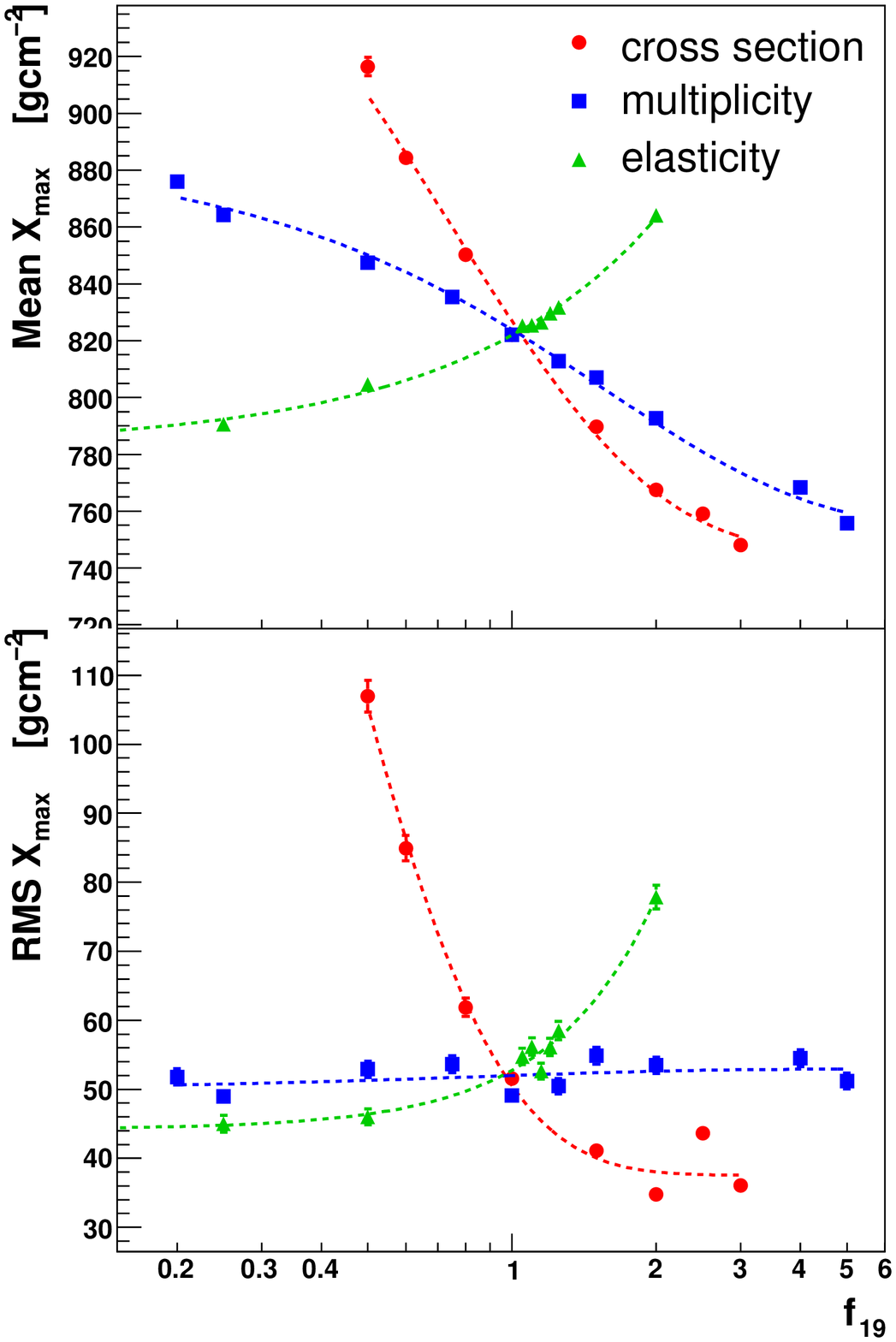}}
    \subfloat[Electrons, $N_{\rm e}$]{\includegraphics[width=.34\linewidth]{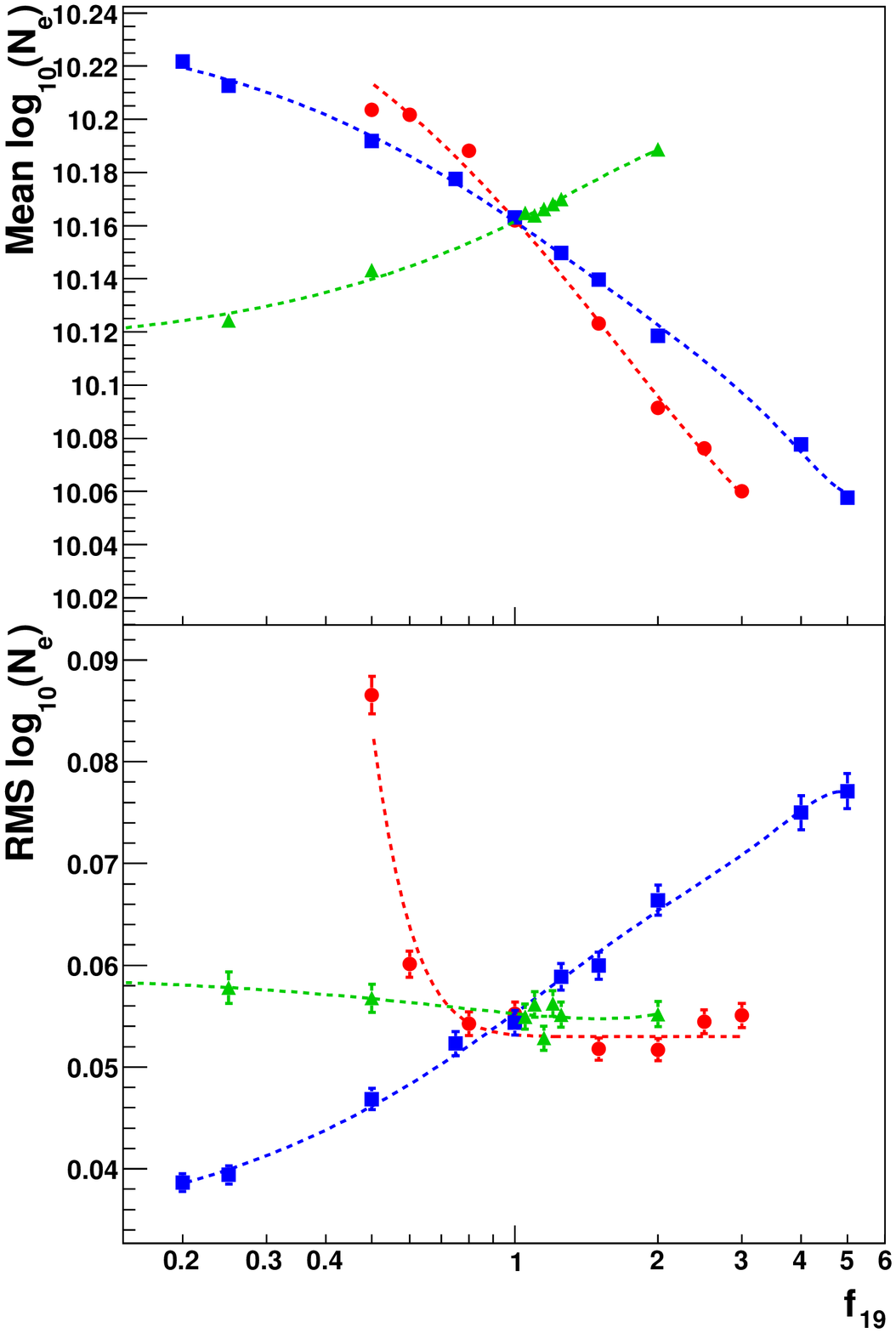}}
    \subfloat[Muons, $N_\mu$]{\includegraphics[width=.34\linewidth]{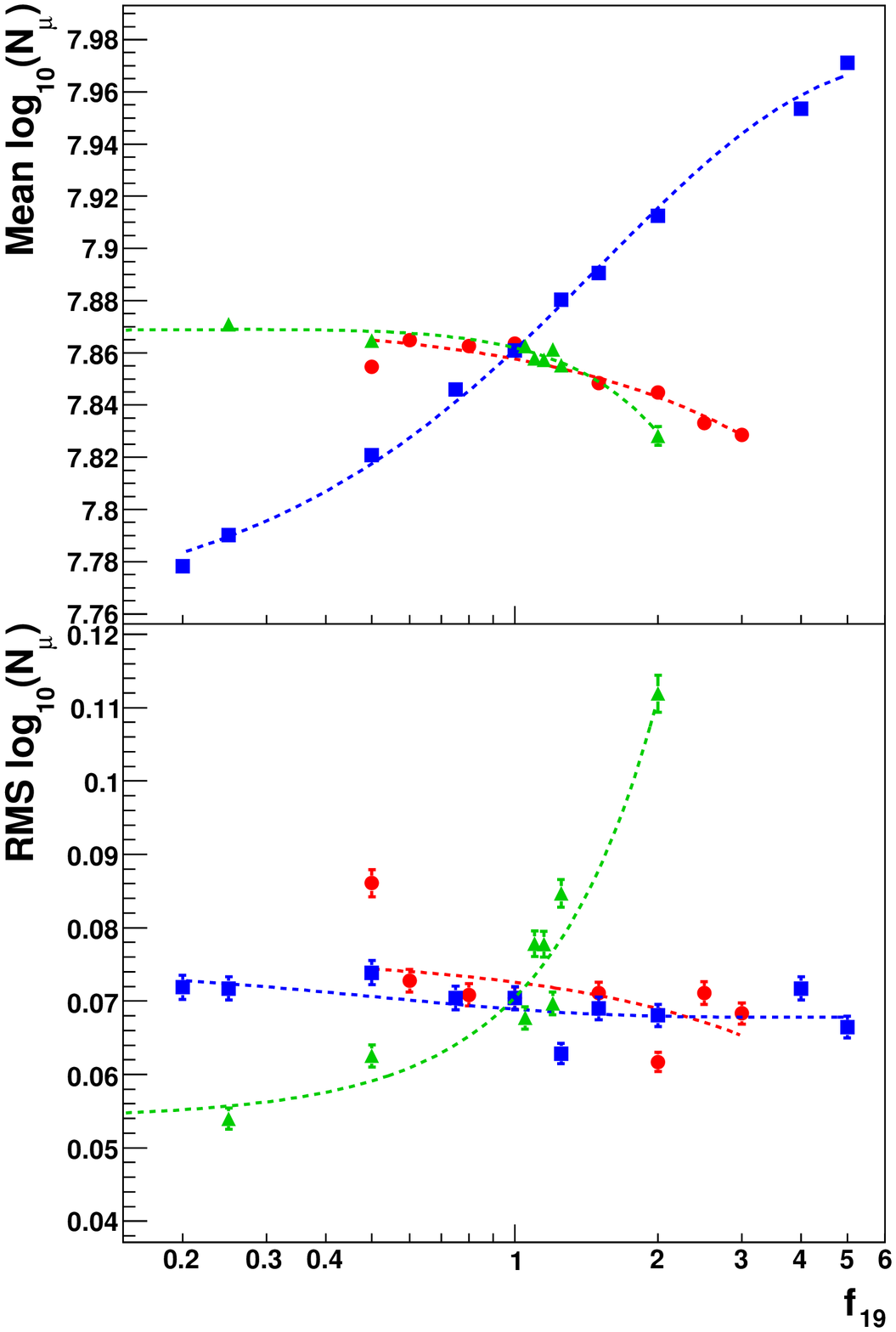}}
  }
  \vspace*{.5cm}
  \caption{Effect of changing interaction characteristics on proton
    induced air showers. Shown is the impact on the observables $X_{\rm
      max}$, $N_{\rm e}$ and $N_\mu$. Each data point is the mean
    value for 1000 simulated air showers at a primary energy of
    $10^{19.5}\,$eV. The lines are just to guide the eye.}
  \label{fig:resultsP}
\end{figure*}

\begin{figure*}[p!]
  \centerline{%
    \subfloat[Shower maximum, $X_{\rm max}$]{\includegraphics[width=.34\linewidth]{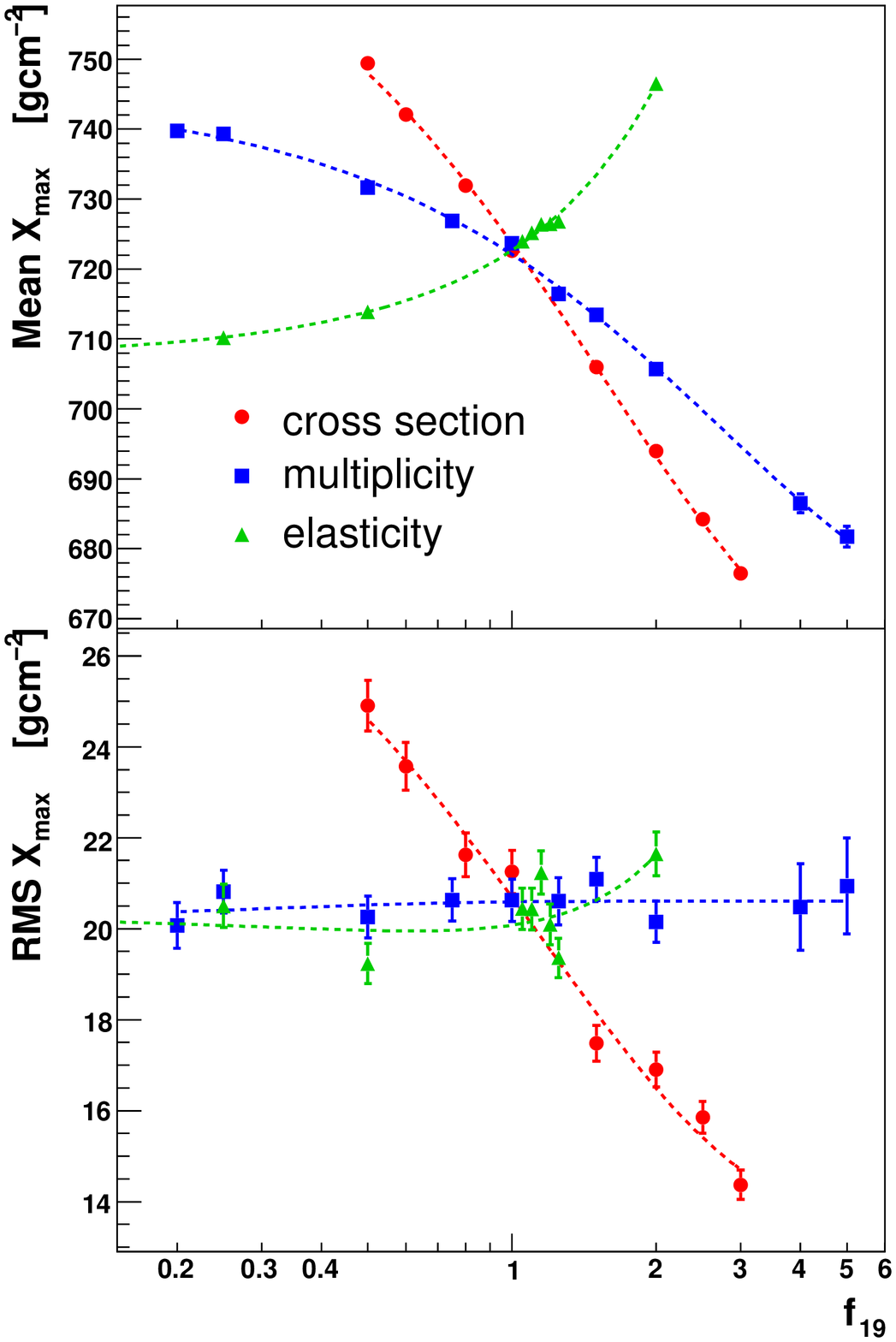}}
    \subfloat[Electrons, $N_{\rm e}$]{\includegraphics[width=.34\linewidth]{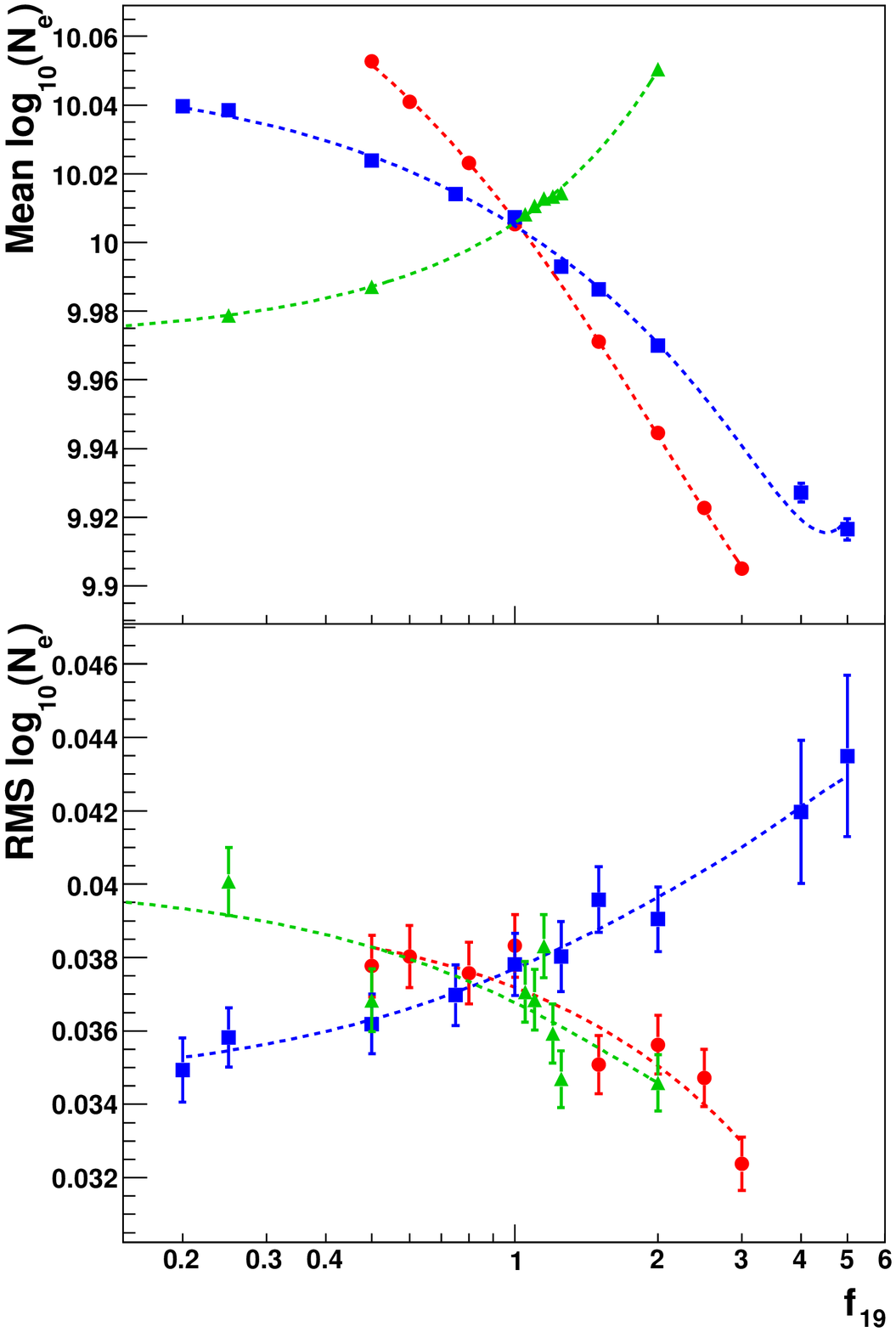}}
    \subfloat[Muons, $N_\mu$]{\includegraphics[width=.34\linewidth]{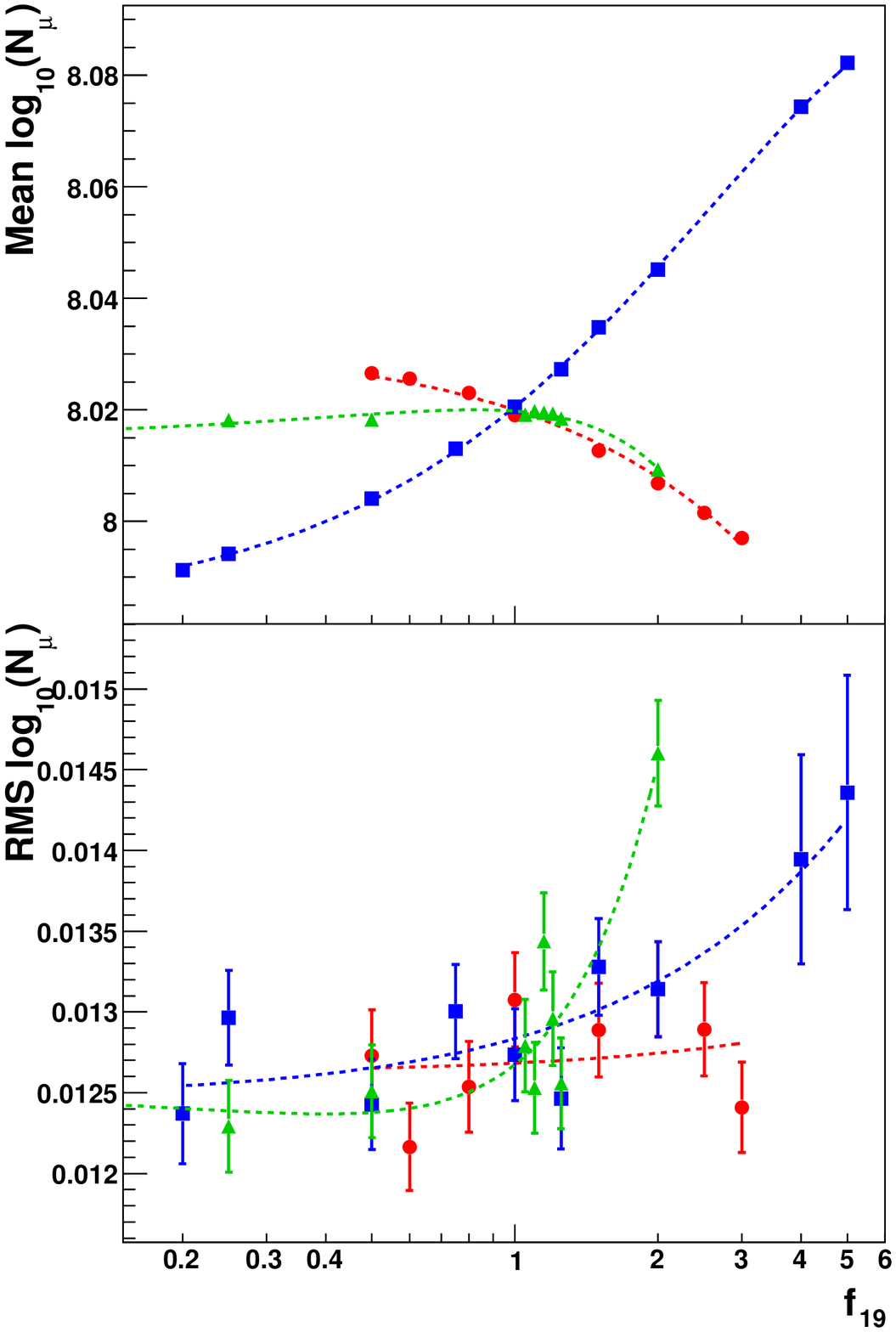}}
  }
  \vspace*{.5cm}
  \caption{Effect of changing interaction characteristics on iron
    induced air showers. Shown is the impact on the observables $X_{\rm
      max}$, $N_{\rm e}$ and $N_\mu$. Each data point is the mean
    value for 1000 simulated air showers at a primary energy of
    $10^{19.5}\,$eV. The lines are just to guide the eye.}
  \label{fig:resultsFe}
\end{figure*}

The results for proton primaries are summarized in
Fig.~\ref{fig:resultsP} and for iron primaries in
Fig.~\ref{fig:resultsFe}.

%

\subsection{Cross section}
A changing cross section has a strong impact on $X_{\rm max}$. Both
the mean as well as the fluctuations are affected. Especially for the
fluctuation, the cross section is much more important than any other
hadronic interaction feature. The effect on the electron number is
related to the changing distance from the shower maximum to the
observation level. Muon are only weakly affected.

For iron primaries the effects are very much reduced. Interestingly the
impact on the mean $X_{\rm max}$ is still very notable, while it changes the 
fluctuations only by up to a few g/cm$^2$.

\subsection{Multiplicity}
The multiplicity shifts the mean value of $X_{\rm max}$ while leaving
its fluctuations almost untouched. The electron number is reduced for
a growing multiplicity, since the shower maximum moves further away
from the detector level. This is also why the fluctuations are
increasing at the same time. The muon number, on the other hand, grows
since it does not depend strongly on the distance to the shower
maximum. This inverse reaction of electron and muon numbers on a
changing multiplicity certainly has interesting implications on
$N_{\rm e}/N_\mu$ unfolding technique, as practiced e.g. by the
KASCADE Collaboration~\cite{Antoni:2005wq}; By changing the
multiplicity, the model predictions move on diagonal lines in the
$N_{\rm e}/N_\mu$-plane.

The same effects are observed on a reduced scale for iron primaries.

\subsection{Elasticity}
The elasticity has an influence on the mean as well as RMS of the
$X_{\rm max}$ distribution. The case of an increased elasticity by
$f_{19}=2$, as we included it in our results, is the most extreme
modification of hadronic interactions that we present; The elasticity
is not expected to rise at ultra-high energies.  Again, the electron
number reacts to the shifting $X_{\rm max}$. A surprisingly strong
effect is observed in the fluctuations of the muon signal.

The latter effect disappears for iron primaries. The impact on $X_{\rm
  max}$ and $N_{\rm e}$ is comparable to the one induced by
multiplicity and cross section.


\section{Summary}
We demonstrate the importance of the extrapolation of hadronic
interaction features from accelerator data to ultra-high energies for
air shower development. For this purpose hadronic interactions are
modified during the air shower simulation process within a customized
version of the \textsc{Conex} program. 

It is found that the resulting impact on air shower observables is
much larger than by just considering the properties of the single
first interaction of the primary cosmic ray particle in the
atmosphere. For example the predicted value of $\langle X_{\rm
  max}\rangle$ for primary iron nuclei changes by up to
$>\unit[30]{g/cm^2}$ while changing the cross section by a factor of
2; Since the mean free path of iron in air itself is only
$\unit[\sim8]{g/cm^2}$ a change by a factor of 2 can only explain $4$
respectively $\unit[16]{g/cm^2}$ of the total impact. The remaining
shift of $\langle X_{\rm max}\rangle$ is in fact originating from a
different air shower development after the first interaction.

\newpage
Currently, the existing uncertainties in interaction physics at cosmic
ray energies prevent an unambiguous interpretation of air shower data
in terms of e.g. the primary cosmic ray mass composition.

The demonstrated sensitivity of standard air shower observables on
hadronic interactions characteristics can be exploited to put
constrains on hadronic interaction physics at energies far above the
LHC. Furthermore, if the composition of the cosmic ray flux in a
specific energy region can be inferred from astrophysical
considerations, existing and future high quality air
experiments~[10\,-\,13]
can be used in order to explore particle physics up to
$\sqrt{s}~\sim\unit[450]{TeV}$. This is possible for proton cosmic ray
primaries but, with somewhat limited sensitivity, also for primary
iron nuclei.


\end{document}